\documentclass[12pt]{article}
\usepackage{epsfig}
\def\be{\begin{equation}}
\def\ee{\end{equation}}
\def\ba{\begin{eqnarray}}
\def\ea{\end{eqnarray}}
\def\ga{\mathrel{\raise.3ex\hbox{$>$\kern-.75em\lower1ex\hbox{$\sim$}}}}
\def\la{\mathrel{\raise.3ex\hbox{$<$\kern-.75em\lower1ex\hbox{$\sim$}}}}

\begin{document}
\baselineskip=16pt
\begin{titlepage}
\begin{center}

\vspace{0.5cm}
{\Large \bf
Time varying $\alpha$ in $N=8$ extended Supergravity
}\\
\vspace{10mm}

Qing-Guo Huang$^{a}$ and Miao Li$^{a}$ \\
\vspace{6mm}
{\footnotesize{\it
  $^a$Institute of Theoretical Physics, Chinese Academy of Sciences,\\
      P.O. Box 2735, Beijing 100080, China \footnote{Email address:
      huangqg@itp.ac.cn,mli@itp.ac.cn}\\}}

\vspace*{5mm}
\normalsize
\smallskip
\medskip
\smallskip
\end{center}
\vskip0.6in
\centerline{\large\bf Abstract}
 {There has been some evidence that the fine structure ``constant" $\alpha$
may vary with time. We point out that this variation can be described 
by a scalar field in some supergravity theory
in our toy model, for instance, the $N=8$
extended supergravity  in four dimensions which can be accommodated in M-theory.}
\vspace*{2mm}
\end{titlepage}

There exists a lot of analysis about the observational
constraints on possible variation of the fine structure constant $\alpha$
\cite{oc} in time. 
A number of absorption systems in the spectra of distant quasars suggests
a smaller value of $\alpha$ in the past, with a favored value of the 
change 
$\frac{\delta \alpha}{\alpha} \cong (-0.72 \pm 0.18) \times {10^{-5}}$
over the red-shift range 
$0.5 \leq z \leq 3.5$ 
\cite{oc1}.
And the analysis of the isotope abundances in the Oklo natural reactor
operated $1.8$ Gyr ago gives a new constraint,
$\frac{|\delta \alpha|}{\alpha} \sim 10^{-7}$
\cite{o}. Some theoretical issues of a varying $\alpha$ are discussed in
\cite{m1,m2, m3}.

There are two scenarios to explain the variation of $\alpha$.
One possibility is that there was a first order phase transition between
the time at which the quasar light was emitted and the present.
But if the Oklo natural reactor confirms the above data which indicates that
$\alpha$ at the red-shift $z \sim 0.13$ is different from its present value,
it is most possible that $\alpha$ varied with time continuously. On the other 
hand, in string or M-theory, the couplings in the effective field theory
depend on the expectation values of some dynamical scalar fields such as the dilation 
and other string moduli, then the coupling ``constants" in general vary with 
time, if they are not trapped in the minimum of a potential in an early time.
Naturally the fine structure constant varies continuously in this scenario.

In a four dimensional effective field theory, the change of the fine structure constant
is controlled by a dynamical scalar field $\phi$, which may be a combination
of several canonically normalized moduli scalars \cite{m1,m2,m3}.
The change in $\alpha$ during the last Hubble time requires that the scalar
field $\phi$ should be extraordinarily light, with a mass comparable to the 
present Hubble scale $H_0 \sim 10^{-33} eV$ \cite{m1}.
It is usually difficult to find a candidate for this scalar field in a fundamental theory.

However it was found that one can describe the present state of 
quasi-exponential expansion of the universe in a broad class of models based on 
four dimensional $N=8$ extended supergravity \cite{l1,l2,l3}.
And it is important that there are scalars whose masses in $N \geq 2$ supergravity 
are quantized in units of the Hubble constant $H_0$ corresponding to DS solutions:
$\frac{m^2}{H_0^2}=-n$, where n are some positive 
integers of the order 1.
The minus sign says that the scalar fields are tachyonic, consistent with
the fact that $\alpha$ becomes larger and larger.
In particular, there is a scalar whose mass square of the scalar field $m^2=-6H^2$ in 
$N=8$ supergravity \cite{l1}.
And we know N=8 supergravity with de Sitter maximim and 
one scalar field
has only non-Abelian gauge fields $SO(3) \times SO(5)$ 
or $SO(4) \times SO(4)$.
But in our real world the supersymmetry must be broken and then 
the Abelian gauge field will appear in our toy model.

In an effective field theory or M-theory, the photon kinetic term reads

\be
f(\phi) F_{\mu\nu} F^{\mu\nu}
\label{cop}
\ee
where $f(\phi)$ is a function of $\phi$.
The most general expansion of the function $\alpha(\phi)$ about
its present day value $\alpha_0=\alpha(\phi_0)$ is
\be
\alpha(\phi)=\alpha_0+\lambda_{\phi} \frac{\delta \phi}{M}+...
\ee
where M is a typical scale over which $\phi$ varies.
The variation of $\alpha$ with $\phi$ is generally written to the
leading order in $\phi$ as
\be
\frac{\delta \alpha}{\alpha} \approx 
\frac{\lambda_{\phi} \delta \phi}{\alpha_0 M}
\label{al}
\ee
where 
$\delta \alpha=\alpha(z)-\alpha_0$,
$\delta \phi=\phi(z)-\phi_0$ and z is cosmological red-shift.

For simplicity, we take $\phi$ to be governed by the
Lagrangian
$
{\cal{L}}=(\partial \phi)^2-m^{2} \phi^2+...$, namely $\phi$ is a
canonical scalar field.
As usual, we assume the scalar field $\phi$ be homogeneous and the 
equation of motion for $\phi$ reads
\be
\ddot{\phi}+3 H_0 \dot{\phi}+ m^2 \phi^2+...=0
\label{em}
\ee
We take $m^2$ to be negative, or simply switch to the notion $m^2
\rightarrow -m^2$ with positive $m^2$.

We consider in general the case when $m^2$ is the same order
as $H^2$, which means the scalar field is not slow rolling.
Firstly we assume the Hubble constant is a constant and
assume $\phi$ has the form as $\phi(t)=\phi_0 e^{i \omega (t-t_0)}$.
Equation (\ref{em}) leads to
\be
\omega=iH(\frac{3}{2} \pm \sqrt{\frac{9}{4}+\frac{m^2}{H^2}})
\ee
Choose the solution with the minus sign corresponding to 
the exponentially growing solution
\be
\phi(z)=\phi(t)=\phi_0 
exp [ H(t-t_0) F(\frac{m^2}{H^2})  ]
=\phi_0 exp[ - z F(\frac{m^2}{H^2})]
\ee
with
\be
F(\frac{m^2}{H^2})=\sqrt{\frac{9}{4}+\frac{m^2}{H^2}}-\frac{3}{2}.
\ee
Since we are interested only in the range $0 \leq z \leq 1$, we
can set $H(t_0-t)=z$. Substitute this result to the equation (\ref{al}),
we find
\be
\frac{\delta \alpha(z)}{\alpha} \approx
\Lambda (exp[ - z F(\frac{m^2}{H^2})]-1)
\label{va}
\ee
here $\Lambda$ is a constant which can be fixed by experiment.
As an example, in $N=8$ supergravity, the mass square
of the scalar field is $-6 H^2$ in order to get a DS solution, so
$m^2=6H^2$ in equation (\ref{va}) and $F(\frac{m^2}{H^2})=1.37$.
Using the data of \cite{oc1}
$\frac{\delta \alpha(z=1)}{\alpha}=-0.7 \times 10^{-5}$ fixes
$\Lambda=9.38 \times 10^{-6}$.
Thus
\be
\frac{\delta \alpha(z)}{\alpha} \approx
9.38 \times 10^{-6} (e^{-1.37 z}-1)
\label{n8}
\ee






Putting $z=0.13$ into formula (\ref{n8}), we get
 $\frac{\delta \alpha(z=0.13)}{\alpha} \approx -1.5 \times
10^{-6} \sim - 10^{-7}$. It is slightly off the constraints of the 
Oklo experiment. To better satisfy all constraints (assuming they
are all reliable), we can increase $m^2$ and decrease $\Lambda$, 
and still keep $m^2$ reasonably close to $H^2$.

In fact the Hubble parameter is not a constant during this stage.
To be more precise, we can numerically solve the equations which Linde et.
suggested in \cite{l3}. We also assume that the universe is 
spatially flat. Thus we have
\be
\ddot{\phi}+3H\dot{\phi}+V,_{\phi}=0
\label{mo}
\ee

\be
H^2=\frac{1}{3} (\rho_M + \frac{1}{2} {\dot{\phi}}^2 + V(\phi) )
\ee

\be
\rho_{total}=\rho_M+ \frac{1}{2} {\dot{\phi}}^2 + V(\phi)
\ee

where we consider the extended $N=8$ supergravity which tells us
$V(\phi)=3 H_0^2 (2-cosh(\sqrt{2} \phi))$ 
and we use the method of  \cite{l3}   
(in units $M_p=1$ and the present Hubble parameter $H_0=1$). 
Here we assume that the Hubble parameter in de Sitter regime approximately 
equals the present Hubble parameter $H_0$. 
In our calculations we shall also assume that initially the matter
energy density $\rho_M$ is much greater than the energy density of the
scalar field $\phi$. Thus the scalar field $\phi$ freezes and its
initial velocity can be set to zero resonablely, since at that moment
the friction term in (\ref{mo})  is very large. 
According to 
equation (\ref{mo}),the scalar field $\phi$ will not change if its 
initial value is zero. In our case we choose the iniatial value
of $\phi$ leading to the dark energy $\Omega_D=0.73$ today \cite{wmap}. 
We show the numerical solution in Figure 1 (here we take
equation (\ref{al}) into account and describe the variation of $\alpha$
with time directly).

\begin{figure}[ht]
\begin{center}

\mbox{\epsfig{file=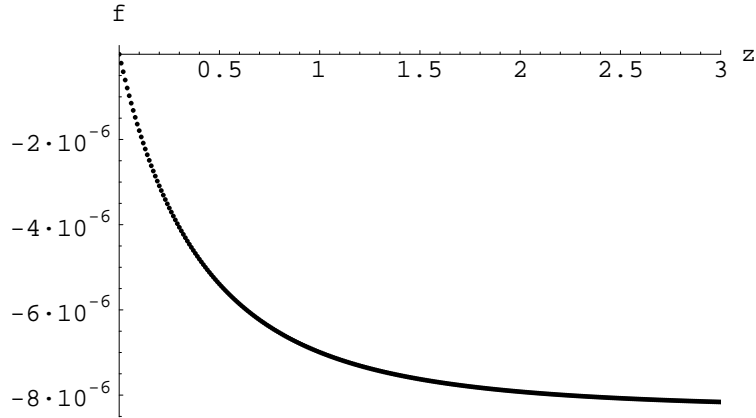,width=10cm}}

\caption {In this Figure $f=\frac{\delta \alpha (z)} {\alpha}$ and z is
the cosmological redshift.}

\end{center}
\end{figure}

In Figure 1 we fit the data 
$\frac{\delta \alpha(z=1)}{\alpha}=-0.7 \times 10^{-5}$. 
According to this figure we can read out  
$\frac{\delta \alpha(z)}{\alpha} \sim -1 \times 10^{-5}$,
which is consistent with \cite{oc1}. And \cite{oc1} also
requires $\frac{\dot \alpha}{\alpha} = (-2.2 \pm 5.1) \times 10^{-16} yr^{-1}$
over the red-shift range $0.6 \leq z \leq 1.0$. 
Using $H_0=71 km \cdot s^{-1} \cdot Mpc^{-1}$ we predict 
$\frac{\dot \alpha}{\alpha}=-5 \times 10^{-16} yr^{-1}$ around $z=1$.

In short,
in this paper we suggest that scalar fields in some supergravity theory
with a mass comparable to the Hubble scale can be a candidate for describing 
the variation of the fine structure constant.
Linde et al. have discussed some possibilities for
describing our universe with a tiny positive cosmological constant 
in extended $N=2,4,8$ supergravity theories,
and suggested that some scalar fields in these supergravity theories can be used to 
describe the present stage of our universe, but not the early universe.

The authors of \cite{m3} pointed out that in an effective field theory
with a cut-off, such as a theory accommodating SUSY breaking, a varying
fine structure constant is always accompanied by a variation in the vacuum
energy or the cosmological constant, which in general is too large. We do not 
know how the resolve this problem.

Other theoretical attempts on explaining a varying fine structure constant can be
found in \cite{ot}.

\textbf{Acknowledgments}

We would are grateful to Yun-Song Piao for useful discussions.


\begin{thebibliography}{99}

\bibitem{oc} T. Damour, F. Dyson, Nucl. Phys.B 480, 37 (1996);\\
P. Sisterna, H. Vucetich, Phys. Rev. D41, 1034 (1990);\\
D. Prestage et al., Phys. Rev. Lett. 74, 3511 (1995);\\
J. Bernstein, L. S. Brown, G. Feinberg, Rev. Mod. Phys. 61, 25 (1989);\\
P. P. Avelino et al., Phys. Rev. D62, 123508 (2000);\\
R. A. Battye et al., Phys. Rev. D63, 044505 (2001);\\
S. Landau, D. H. Harari, M. Zaldarriaga, Phys. Rev. D63, 083505 (2001);\\
J. P. Uzan, Rev. Mod. Phys. 75 (2003) 403.

\bibitem{oc1} J. K. Webb et al., Phys. Rev. Lett. 87, 091301 (2001),
astro-ph/0012539.\\
J.K.Webb et al.,Phys.Rev.Lett.82,884(1999).

\bibitem{o} Y. Fujii, A. Iwamoto, T. Fukahori, T. Ohnuki, M. Nakagawa,
H. Hikida, Y. Oura, P. Moller, Nucl. Phys. B573, 377 (2000),\\
Y.Fujii,Astrophys.Space Sci,283(2003)559-564(qr-gc/0212017).


\bibitem{m1} G. Dvali, M. Zaldarriaga, Phys. Rev. Lett.88, 091303 (2002).

\bibitem{m2} T. Chiba, K. Kohri, Prog. Theor. Phys. 107 (2002) 631-636.

\bibitem{m3} T. Banks, M. Dine, M. Douglas, Phys. Rev. Lett.88, 131301
(2002).

\bibitem{l1} R. Kallosh, A. Linde, S. Prokushkin, M. Shmakova, Phys.
Rev. D65, 105016 (2002).

\bibitem{l2} R. Kallosh, A. Linde, astro-ph/0301087.

\bibitem{l3} R.Kallosh, hep-th/0205315;\\
R. Kallosh, A. Linde, S. Prokushkin, M. Shmakova, hep-th/0208156;\\
R. Kallosh, A. Linde, hep-th/0208157;\\
P. Fre, M. Trigiante, A. Van Proeyen, hep-th/0205119;\\
M.Gutperle,R.Kallosh,A.Linde,hep-th/0304225.

\bibitem{wmap} C.L.Bennett et al.,astro-ph/0302207.

\bibitem{ot} S. Das, G. Kunstatter, hep-th/0212334;\\
J. D. Barrow, D. F. Mota, gr-qc/0212032;\\
J. D. Barrow, gr-qc/0211074;\\
F. P. Correia, M. G. Schmidt, Z. Tavartkiladze, hep-ph/0211122;\\
J. D. Bekenstein, Phys. Rev. D66, 123514 (2002).


\end{thebibliography}
\end{document}